\documentclass[prd,aps,twocolumn,preprintnumbers, showpacs, nofootinbib,superscriptaddress,notitlepage]{revtex4-1}

\usepackage{amsmath,graphicx,epsfig,amssymb,dsfont,mathtools}
\usepackage[usenames]{color}
\usepackage{ulem} 
\usepackage{bigstrut}
\usepackage{slashed}
\usepackage{multirow}
\usepackage{subfigure}

\allowdisplaybreaks

\begin{document}
	
\title{Heavy flavor conserved semi-leptonic decay of $B_s$ in the covariant light-front approach}
	
\author{Yu-Ji Shi}
\email{shiyuji@ecust.edu.cn}
\affiliation{School of Physics, East China University of Science and Technology, Shanghai 200237, China }
\affiliation{INPAC, Key Laboratory for Particle Astrophysics and Cosmology (MOE), Shanghai Key Laboratory for Particle Physics and Cosmology,
School of Physics and Astronomy, Shanghai Jiao Tong University, Shanghai 200240, China}

\author{Zhi-Peng Xing}
\email{zpxing@sjtu.edu.cn}
\affiliation{Tsung-Dao Lee Institute, Shanghai Jiao Tong University, Shanghai 200240, China}


\begin{abstract}
We study the heavy flavor conserved semi-leptonic decay $B_s\to B\ell\nu$ in the covariant light front approach.  The covariant light front quark model is used to calculate the transition form factors of $B_s\to B^{(*)}$ as well as $D_s\to D$, which are consistent with the leading power predictions from the heavy quark symmetry.  The angular distribution analysis  on the $B_s\to B^{(*)}l\bar\nu$ decay is performed by investigating the forward-backward asymmetry of the lepton. We also study the angular distribution of $B_{s}\to B^{*}(\to B \gamma)l\bar\nu$ decay both through the lepton forward-backward asymmetry and the azimuth angle.  The branching fractions of $B_s\to Bl\bar\nu$ and $B_s\to B^{*}l\bar\nu$ are  at the order $10^{-8}$ and $10^{-9}$, respectively. The number of  $B_s\to B l\bar\nu$ events is estimated to be $1.76$. The branching fraction of $B_{s}\to B^{*}(\to B \gamma)l\bar\nu$ is at the order $10^{-11}$, which is calculated by introducing Breit-Wigner distribution for the intermediate $B^*$.
\end{abstract}

\maketitle

\section{Introduction}
Heavy flavor physics is an important research area in particle physics due to its crucial role in precisely testing the Standard Model (SM) and searching for new physics (NP). Generally, highly suppressed decay processes in the SM such as the flavor-changing neutral currents (FCNC) processes are widely used as sensitive probes for NP due to their tiny branching fractions. However, besides the FCNC processes, heavy flavor conserved (HFC) processes also have tiny branching fractions, so they enjoy the same advantage for probing NP.  Recently, the HFC processes have received increasing attention from both experimentalists~\cite{LHCb:2015une,LHCb:2020gge} and theorists~\cite{Voloshin:2019ngb,Niu:2021qcc,Cheng:2022kea,Cheng:2022jbr,Liu:2022mxv}.  Particularly, on the theoretical side, the small recoil in HFC processes will effectively constrain the form factor for the transition matrix element. The limited phase space will enhance the amplitudes suppressed by lepton mass in semi-leptonic decays, which is also referred to as helicity suppressed amplitudes. As discussed in Ref.~\cite{Datta:2017aue},  when the NP Hamiltonian with $(S-P)$ and tensor current are introduced,  helicity suppressed amplitudes will appear due to the interference of the NP and SM currents. These amplitudes are sensitive to the NP effects since the NP contribution in positive definite term is highly suppressed by the NP Wilson coefficients. Therefore, these effects are very helpful for NP studies and unique in HFC processes.

The HFC decays of $B_s$ have highly suppressed branching fractions due to the limited phase space of the $s\to u$ transition. This feature makes them ideal platforms for detecting NP, but they have so far not received much attention due to the difficulty of experimental measurement. However, nowadays the abundant production of $B_s$ in the Belle experiment will improve the situation.
As an electron-positron collider, the Belle detector is currently capable of collecting $711\ {\rm fb}^{-1}$ at the energy arround the $\Upsilon(4S)$ mass, and $121.4\ {\rm fb}^{-1}$ at the energy arround the $\Upsilon(5S)$ mass~\cite{Dubey:2022lyq}. $\Upsilon(5S)$ can decay into $B_s^{(*)}$ in three ways: $\Upsilon(5S)\to B^{0*}_s\bar{B}^{0*}_s$, $B^{0*}_s\bar{B}^{0} (B^{0}_s\bar{B}^{0*}_s$) and $B^{0}_s\bar{B}^{0}_s$ with branching fractions 0.87, 0.073 and 0.057, respectively \cite{Belle:2012tsw}. Therefore, the data samples of $B_s$ generated by the Belle experiment are sufficient for precise measurements of various HFC $B_s$ decays.
  
In this work, we will study the semi-leptonic decay $B_s\to B^{(*)}l\bar\nu$, which is the simplest HFC process of $B_s$, using the covariant light-front quark model (LFQM). The traditional LFQM model \cite{Terentev:1976jk,Berestetsky:1977zk,Jaus:1989au,Jaus:1989av} was originally developed based on the  light-front formalism \cite{Brodsky:1997de}, which has been widely applied for calculating non-perturbative quantities in terms of hadrons.  However, the traditional LFQM loses Lorentz invariance and suffers from the so-called zero mode problem \cite{Choi:1998nf}. The covariant LFQM  was first proposed in Refs. \cite{Jaus:1999zv,Choi:1998nf,Cheng:1997au} with the use of covariant Bethe-Salpeter (BS) approach~\cite{Salpeter:1951sz,Salpeter:1952ib}, and was then widely applied for meson decays \cite{Jaus:2002sv,Cheng:2003sm,Bakker:2000pk,Bakker:2002mt,Bakker:2003up,Choi:2005fj,Choi:2013mda,Chang:2018zjq,Chang:2019mmh,Zhang:2023ypl}. Recently, the LFQM has also been successfully applied for heavy baryon decays with the help of the quark-diquark picture \cite{Ke:2007tg,Wei:2009np,Ke:2012wa,Zhao:2018mrg,Wang:2017mqp,Wang:2022ias,Liu:2022mxv,Zhao:2022vfr,Zhang:2022bvl,Geng:2022xpn,Geng:2020fng,Hsiao:2020gtc,Liu:2023zvh} or the three quark picture \cite{Zhao:2023yuk}. Therefore, the covariant LFQM must be reliable for our study of $B_s\to B^{(*)}l\bar\nu$ decays.

This paper is organized as follows. In section II, we will present the calculation of the transition form factors in the  framework of covariant LFQM. The corresponding numerical calculation and the angular distribution analysis will be performed in section III.  A brief summary is given in the last section.

\section{$B_{s}\to B^{(*)}\bar{l}\nu$ decay in covariant light-front quark model}
In this section, we introduce the covariant calculation for the transition form factors of $B_s\to B^{(*)}$ using LFQM. The effective electroweak Hamiltonian  for the $B_{s}\to B^{*}\bar{l}\nu$ decay reads as
\begin{equation}
{\cal H}_{{\rm eff}}=\frac{G_{F}}{\sqrt{2}}V_{us}[\bar{u}\gamma_{\mu}(1-\gamma_{5})s][\bar{l}\gamma^{\mu}(1-\gamma_{5})\nu],\label{eq:Hamilt}
\end{equation}
where $G_{F}$ is the  Fermi constant. With this Hamiltonian, the leptonic part of the decay amplitude can be computed straightforwardly, while the hadronic part are parameterized by the transition form factors of $B_s\to B^{(*)}$. The corresponding form factors induced by the vector and axial-vector currents are defined as
\begin{align}
&\langle B(P^{\prime})|\bar{u}\gamma_{\mu}s|B_{s}(P)\rangle \nonumber \\
=&  \left({\bar P}_{\mu}-\frac{m_{B_{s}}^{2}-m_{B}^{2}}{q^{2}}q_{\mu}\right)F_{1}(q^{2}) +\frac{m_{B_{s}}^{2}-m_{B}^{2}}{q^{2}}q_{\mu}F_{0}(q^{2}),\nonumber \\
&\langle B^*(P^{\prime},\varepsilon^{\prime})|\bar{u}\gamma_{\mu}s|B_{s}(P)\rangle\nonumber \\
  =&  \ \frac{2V(q^{2})}{m_{B_{s}}+m_{B^{*}}}\,\epsilon_{\mu\nu\alpha\beta}\varepsilon^{\prime}{}^{*\nu}P^{\alpha}P^{\prime\beta},\nonumber \\
&\langle B^*(P^{\prime},\varepsilon^{\prime})|\bar{u}\gamma_{\mu}\gamma_{5} s|B_{s}(P)\rangle\nonumber \\
  =& \  2i m_{B^{*}}\,\frac{\varepsilon^{\prime*}\cdot q}{q^{2}}\,q_{\mu}A_{0}(q^{2})\nonumber \\
 &+i(m_{B_{s}}+m_{B^{*}})A_{1}(q^{2})\left[\varepsilon_{\mu}^{\prime\prime*}  - \frac{\varepsilon^{\prime*}\cdot q}{q^2}q_\mu\right] \nonumber \\
 &    -i\frac{\varepsilon^{\prime*}\cdot {\bar P}}{m_{B_{s}}+m_{B^{*}}}\,A_{2}(q^{2})\left[{\bar P}_{\mu}- \frac{m_{B}^2-m_{B^{*}}^2}{q^2} q_\mu \right],\label{eq:formfactors}
\end{align}
where $q=P-P^{\prime}$, ${\bar P}=P+P^{\prime}$ and the Levi-Civita tensor is defined as $\varepsilon_{0123}=1$. 

In the LFQM, the light-front decomposition for a four-momentum vector reads as:  $P=(P^-,P^+,{\vec P}_{\perp})$
with $P^{\pm}=P^{0}\pm P^{3}$, and thus $P^{2}=P^{+}P^{-}-{\vec P}_{\perp}^{2}$. The constituent quark momentums of the initial (final) meson are denoted as $p_{1}^{(\prime)}, p_2^{(\prime)}$. Note that $P^{(\prime)}$ and $p_{1}^{(\prime)}, p_2^{(\prime)}$ cannot be on shell simultaneously. In the covariant LFQM, $P^{(\prime)}$ is on-shell, namely $P^{(\prime)2}=M^{(\prime)2}$ with $M^{(\prime)}$ being the physical mass of the initial (final) meson. However, $p_{1}^{(\prime)}, p_2^{(\prime)}$ are off-shell due to the four-momentum conservation $P^{(\prime)}=p_{1}^{(\prime)}+p_{2}^{(\prime)}$. This situation is different from that in the traditional LFQM, where $p_{1}^{(\prime)}, p_2^{(\prime)}$ are set on shell and only the $+, \perp$ components of the momentums are conserved.
The constituent quark momentums can be expressed by the internal variables $(x_{i}^{(\prime)},{\vec k}_{\perp}^{(\prime)})$ as:
\begin{equation}
p_{1,2}^{(\prime)+}=x_{1,2}^{(\prime)}P^{(\prime)+},\quad {\vec p}_{1,2\perp}^{(\prime)}=x_{1,2}^{(\prime)}{\vec P}_{\perp}^{(\prime)}\pm {\vec k}_{\perp}^{(\prime)}.
\end{equation}
It is convenient to define several internal quantities during the LFQM calculation:
\begin{align}
k_{z}^{(\prime)}  &=  \frac{x_{2}^{(\prime)}M_{0}^{(\prime)}}{2}-\frac{m_{2}^{(\prime)2}+{\vec p}_{\perp}^{(\prime)2}}{2x_{2}M_{0}^{(\prime)}},\nonumber\\
e_{i}^{(\prime)} & = \sqrt{m_{i}^{(\prime)2}+{\vec p}_{\perp}^{(\prime)2}+k_{z}^{(\prime)2}},\nonumber\\
M_{0}^{(\prime)2} & = (e_{1}^{(\prime)}+e_{2}^{(\prime)})^{2}=\frac{{\vec k}_{\perp}^{(\prime)2}+m_{1}^{(\prime)2}}{x_{1}^{(\prime)}}+\frac{{\vec k}_{\perp}^{(\prime)2}+m_{2}^{(\prime)2}}{x_{2}^{(\prime)}},\nonumber\\
\widetilde{M}_{0}^{(\prime)}  &=  \sqrt{M_{0}^{(\prime)2}-(m_{1}^{(\prime)}-m_{2}^{(\prime)})^{2}},
\end{align}
where $i=1,2$, $m_i^{(\prime)}$ are the constituent quark masses, $M_{0}^{(\prime)}$ is the kinetic invariant mass of the initial (final) meson, $k_{z}^{(\prime)}$ is the relative momentum of the constituent quark in the $z$-direction and $e_{i}^{(\prime)}$ is the energy of quark $i$.

\begin{figure*}[htp]
\includegraphics[scale=0.4]{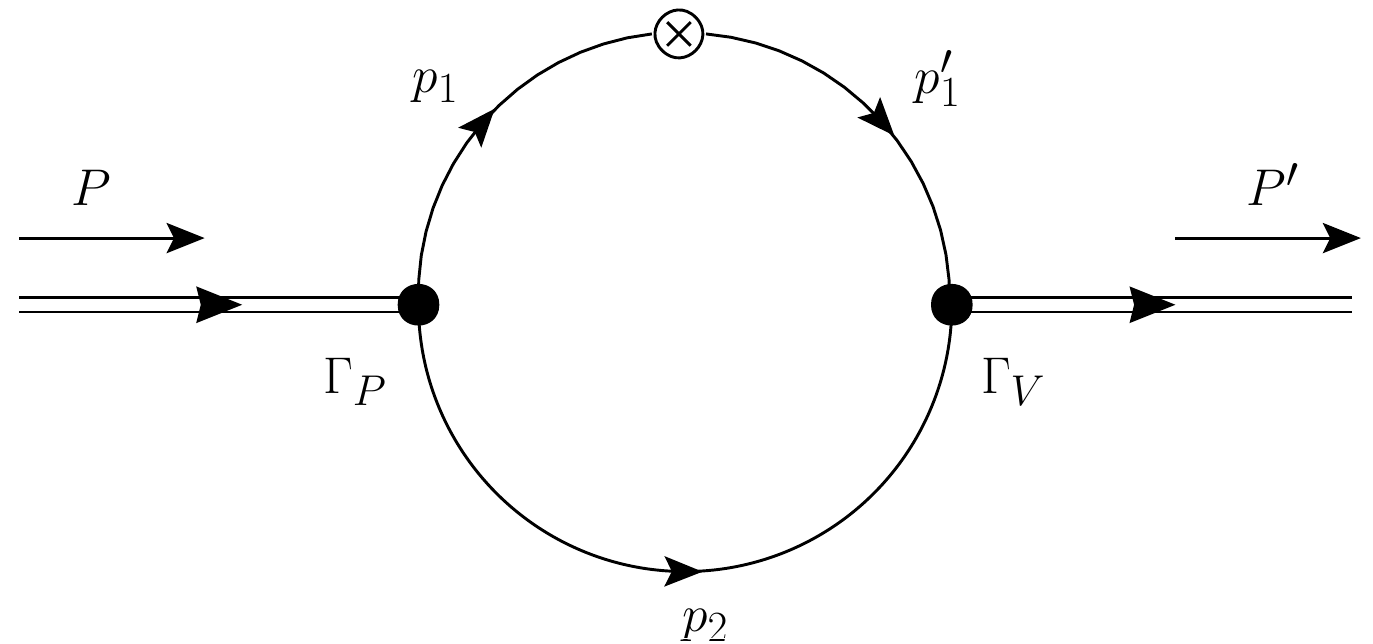}
\caption{Feynman diagram for the calculation of $B_s\to B^{(*)}$ transition form factors, where the
cross symbol denotes the  transition current and the black dots denote the bound state vertexes.}
\label{fig:feynLFQM}
\end{figure*}
The calculation of meson transition matrix elements in the covariant LFQM is equivalent to calculating the amplitude of the Feynman diagram shown in Fig.\ref{fig:feynLFQM} , which can be expressed as 
\begin{equation}
{\cal A}_{\mu}^{{\cal M}{\cal M}^{\prime}}=\frac{N_{c}}{(2\pi)^{4}}\int d^{4}p_{2}\frac{H_{{\cal M}}H_{{\cal M}^{\prime}}}{N_{1}N_{1}^{\prime}N_{2}}i S_{\mu}^{{\cal M}{\cal M}^{\prime}},\label{eq:ampLFQM}
\end{equation}
where $N_c=3$, ${\cal M}$ and ${\cal M}^{\prime}$ denote the initial and final mesons respectively, $H_{{\cal M}},\ H_{{\cal M}^{\prime}}$ are the bound-state vertex functions. The $S_{\mu}^{{\cal M}{\cal M}^{\prime}}$ and the denominators are
\begin{widetext}
\begin{align}
S_{\mu}^{{\cal M}{\cal M}^{\prime}}& = {\rm Tr}[(i\gamma^0\Gamma_{{\cal M}^{\prime}}^{\dagger}\gamma^0)(\slashed p_{1}^{\prime}+m_{1}^{\prime})\Gamma_{\mu}(\slashed p_{1}+m_{1})i\Gamma_{{\cal M}}(-\slashed p_{2}+m_{2})],\nonumber \\
N_{1}^{(\prime)} & = p_{1}^{(\prime)2}-m_{1}^{(\prime)2}+i\epsilon,
\;\;\;
N_{2}   =   p_{2}^{2}-m_{2}^{2}+i\epsilon,
\end{align}
\end{widetext}
where $i\Gamma_{{\cal M}^{(\prime)}}$ is the vertex operators. The vertex operators for pseudoscalar and vector mesons are 
\begin{align}
i \Gamma_{P}=-i \gamma_{5},\ \ \ \ i \Gamma_{V}=i\left[\gamma^{\mu}-\frac{\left(k_{1}-k_{2}\right)^{\mu}}{w_V}\right]
\end{align}
with $w_V=M_0^{\prime}+m_1^{\prime}+m_2$. 

Here we choose a suitable frame where $q^{+}=0$. The $p_{2}^{-}$
integration in Eq.~(\ref{eq:ampLFQM}) picks up the pole $p_{2}=\hat{p}_{2}=[({{\vec p}_{2\perp}^{2}+m_{2}^{2}})/{p_{2}^{+}},p_{2}^{+},{\vec p}_{2\perp}]$
which leads to
\begin{eqnarray}
&&N_{1}^{(\prime)}  \to \hat{N}_{1}^{(\prime)}=x_{1}(M^{(\prime)2}-M_{0}^{(\prime)2}),\nonumber \\
&&H_{{\cal M}}H_{{\cal M}^{\prime}}S_{\mu}^{{\cal M}{\cal M}^{\prime}} \to h_{{\cal M}}h_{{\cal M}^{\prime}}{\hat S}_{\mu}^{{\cal M}{\cal M}^{\prime}},\nonumber \\
&&\int d^{4}p_{2}\frac{1}{N_{1}N_{1}^{\prime}N_{2}}   \to -i \pi\int dx_{2}d^2 {\vec k}_{\perp}\frac{1}{x_2 {\hat N}_{1}{\hat N}_{1}^{\prime}}.
\end{eqnarray}
where ${\hat S}_{\mu}^{{\cal M}{\cal M}^{\prime}}$ means that all the $p_i^{(\prime)}$ appearing in $S_{\mu}^{{\cal M}{\cal M}^{\prime}}$ are replaced by ${\hat p}_i^{(\prime)}$. The condition $q^{+}=0$ leads to $P^+=P^{\prime +},\ p_1^+=p_1^{\prime +}$ and so that $x_{1}=x_{1}^{\prime}$. Using $p_{2\perp}=p_{2\perp}^{\prime}+q_{\perp}$, one can obtain ${\vec k}_{\perp}^{\prime}={\vec k}_{\perp}-x_2 q_{\perp}$.
The explicit form  of $h_{P}^{\prime}$ was derived in Refs.~\cite{Jaus:1999zv,Cheng:2003sm} as
\begin{equation}
h_{{\cal M}^{(\prime)}}=(M^{(\prime)2}-M_{0}^{(\prime)2})\sqrt{\frac{x_{1}^{(\prime)}x_{2}}{N_{c}}}\frac{1}{\sqrt{2}\widetilde{M}_{0}^{(\prime)}}\varphi^{(\prime)},
\end{equation} 
where $\varphi^{(\prime)}$ is the light-front momentum distribution
amplitude for a $s$-wave meson. Generally, it is taken as a Gaussian-type
wave function \cite{Jaus:1999zv,Cheng:2003sm}:
\begin{align}
\varphi^{(\prime)}(x_{2},{\vec k}_{\perp}^{(\prime)})=\ &4\left(\frac{\pi}{\beta^{(\prime)2}}\right)^{3/4}\sqrt{\frac{dk_{z}^{(\prime)}}{dx_{2}}}\nonumber\\
&\times\exp\left(-\frac{k_{z}^{(\prime)2}+{\vec k}_{\perp}^{(\prime)2}}{2\beta^{(\prime)2}}\right),
\end{align}
where $\beta^{(\prime)}$ is a parameter of order $\Lambda_{QCD}$. As shown in Refs.~\cite{Jaus:1999zv,Cheng:2003sm}, the covariant LFQM suffers from the so-called zero mode contribution. In these literatures a systematic program was derived to include such zero mode contribution, which is realized by the following replacements:
\begin{align}
&\hat{p}_{1\mu}  \to  P_{\mu}A_{1}^{(1)}+q_{\mu}A_{2}^{(1)}, \;\;\;
\hat{N}_{2}   \to  Z_{2}, \nonumber \\
&\hat{p}_{1\mu}\hat{N}_{2}  \to  q_{\mu}\left[A_{2}^{(1)}Z_{2}+\frac{q\cdot P}{q^{2}}A_{1}^{(2)}\right].
\end{align}
The expressions of $A_{i}^{(j)}, Z_2$ are
\begin{align}
A_{1}^{(1)}=&\frac{x_{1}}{2}, \quad A_{2}^{(1)}=A_{1}^{(1)}-\frac{{\vec k}_{\perp} \cdot {\vec q}_{\perp}}{q^{2}}, \nonumber\\
A_{1}^{(2)}=&-{\vec k}_{\perp}^2-\frac{\left({\vec k}_{\perp} \cdot {\vec q}_{\perp}\right)^{2}}{q^{2}}\nonumber\\
Z_{2}=&\hat{N}_{1}+m_{1}^{2}-m_{2}^{2}+\left(1-2 x_{1}\right) M^{2}\nonumber\\
&+\left(q^{2}+q \cdot P\right) \frac{{\vec k}_{\perp} \cdot {\vec q}_{\perp}}{q^{2}}.
\end{align}
The analytical results of the from factors can be found in the Appendix~\ref{ExpressFF} and are also given in Refs~\cite{Jaus:1999zv,Cheng:2003sm}, where an alternative parameterization of the form factors as shown in Eq.~(\ref{eq:formfactors2}) is used.

\section{Angular distributions}
\subsection{Angular Distribution of $B_{s}\to B^{(*)}l\bar\nu$}
We firstly perform an angular distribution analysis for the three-body decay $B_{s}\to B^{(*)}l\bar\nu$. Introducing an auxiliary vector particle $V$ with momentum $q=p_l+p_{\bar\nu}$ and polarization vector $\varepsilon_{\mu}^V$, as well as using the following decomposition formula: 
\begin{align}
g_{\mu\nu}=\varepsilon_{\mu}^{V*}(t)\varepsilon_{\nu}^{V}(t)-\sum_{\lambda=\pm,0}\varepsilon_{\mu}^{V*}(\lambda)\varepsilon_{\nu}^{V}(\lambda),
\end{align}
we can express the helicity amplitudes of $B_{s}\to B^{(*)}l\bar\nu$ induced by the Hamiltonian in Eq.~(\ref{eq:Hamilt})  as
\begin{align}
{\cal A}_{B_{s}\to B}^{h_l}=&\frac{G_F}{\sqrt{2}}V_{us}\Big[A_t B^{h_l}_t-\sum_{\lambda=\pm,0}A_{\lambda} B^{h_l}_{\lambda}\Big],\nonumber\\
{\cal A}_{B_{s}\to {B^*}}^{h_R,h_l}=&\frac{G_F}{\sqrt{2}}V_{us}\Big[A_t^{h_R} B^{h_l}_t-\sum_{\lambda=\pm,0}A_{\lambda}^{h_R} B^{h_l}_{\lambda}\Big],
\end{align}
where the hadronic amplitudes $A_{h_V},A_{h_V}^{h_R}$ and the leptonic amplitudes $B^{h_l}_{h_V}$ are defined as
\begin{align}
A_{h_V}=& \langle B|\bar{u}\gamma^{\mu}(1-\gamma_5)s|B_{s}\rangle \varepsilon_{\mu}^{V*}(h_V),\nonumber\\
A^{h_R}_{h_V}=& \langle B^*(h_R)|\bar{u}\gamma^{\mu}(1-\gamma_5)s|B_{s}\rangle \varepsilon_{\mu}^{V*}(h_V),\nonumber\\
B^{h_l}_{h_V}=&\langle l(h_l) \bar \nu|\bar{l}\gamma^{\mu}(1-\gamma_5)\nu|0\rangle \varepsilon_{\mu}^{V}(h_V),\label{eq:hadrlepAmps}
\end{align}
with $h_V=t,0,\pm$ being the helicity of the auxiliary vector particle. The analytical expressions of $A_{h_V},A_{h_V}^{h_R}$ and $B^{h_l}_{h_V}$ can be found in the Appendix

In the rest frame of $B_{s}$, we let the lepton pair and the $B^{(*)}$ meson move along and against the z-axis, respectively. The angular between the  moving direction of the lepton $l$ and the z-axis is denoted as $\theta_l$. The angular distribution for the differential decay width of $B_{s}\to B^{(*)}l\bar\nu$ can be expressed as:
\begin{align}
\frac{d \Gamma_{B_{s}\to B^{(*)}l\bar\nu}}{d q^2 d {\rm cos} \theta_l}=&\frac{G_F^2 |V_{us}|^2 \beta_l \sqrt{\lambda_{B_s}^{(*)}}}{64(2\pi)^3 m_{B_s}^3}\nonumber\\
\times&\Big[L_1^{(*)}+L_2^{(*)} \ {\rm cos}\theta_l+L_3^{(*)}\  {\rm cos}^2\theta_l\Big],
\end{align}
where $\beta_l=1-{\hat m}_l^2$, ${\hat m}_l=m_l/\sqrt{q^2}$ and $\lambda_{B_s}^{(*)}=[m_{B_s}^2-(m_{B^{(*)}}+\sqrt{q^2})^2][m_{B_s}^2-(m_{B^{(*)}}-\sqrt{q^2})^2]$. The coefficients $L_i^{(*)}$ are functions of $q^2=(p_l+p_{\bar\nu})^2$ and their analytical expressions in terms of the helicity amplitudes are
\begin{align}
L_1=&2 \beta_l q^2 \left[|A_0|^2 + |A_t|^2 {\hat m}_l^2\right],\nonumber\\
L_2=&0,\nonumber\\
L_3=&-2 \beta_l^2 q^2 |A_0|^2
\end{align}
for the $B_{s}\to B l\bar\nu$ decay and 
\begin{align}
L_1^{*}=&\beta_l q^2 \left[2|A^0_0|^2 + 2|A^0_t|^2 {\hat m}_l^2\right.\nonumber\\
&\left.+(1+{\hat m}_l^2)(|A_{\parallel}|^2+|A_{\perp}|^2)\right],\nonumber\\
L_2^{*}=&-4\beta_l q^2 \left[{\rm Re}(A^0_0 A^{0*}_t){\hat m}_l^2+{\rm Re}(A_{\parallel} A_{\perp}^*)\right],\nonumber\\
L_3^{*}=&-\beta_l^2 q^2\left[2|A^0_0|^2 -|A_{\parallel}|^2-|A_{\perp}|^2\right].
\end{align}
for the $B_{s}\to B^{*} l\bar\nu$ decay. Here we have made redefinitions: $A_{\parallel}=(A^+_+ + A^-_-)/\sqrt{2}$ and $A_{\perp}=(A^+_+ - A^-_-)/\sqrt{2}$. The normalized forward–backward (FB) asymmetry for $\theta_l$ is defined as
\begin{align}
A^{3,l}_{FB}=&\frac{\left(\int^1_0-\int^0_{-1}\right)d {\rm cos} \theta_l \frac{d \Gamma_{B_{s}\to B^* l\bar\nu}}{d q^2 d {\rm cos} \theta_l}}{\int^1_{-1} d {\rm cos} \theta_l \frac{d \Gamma_{B_{s}\to B^* l\bar\nu}}{d q^2 d {\rm cos} \theta_l}}\nonumber\\
=&\frac{3 L_2^{(*)}}{6 L_1^{(*)}+2 L_3^{(*)}},\label{eq:FBasymBstoBstar}
\end{align}
which vanishes for the  $B_{s}\to B l\bar\nu$ decay.

\subsection{Angular Distribution of $B_{s}\to B^{*}(\to B \gamma)l\bar\nu$}
\begin{figure*}[htp]
\includegraphics[scale=0.4]{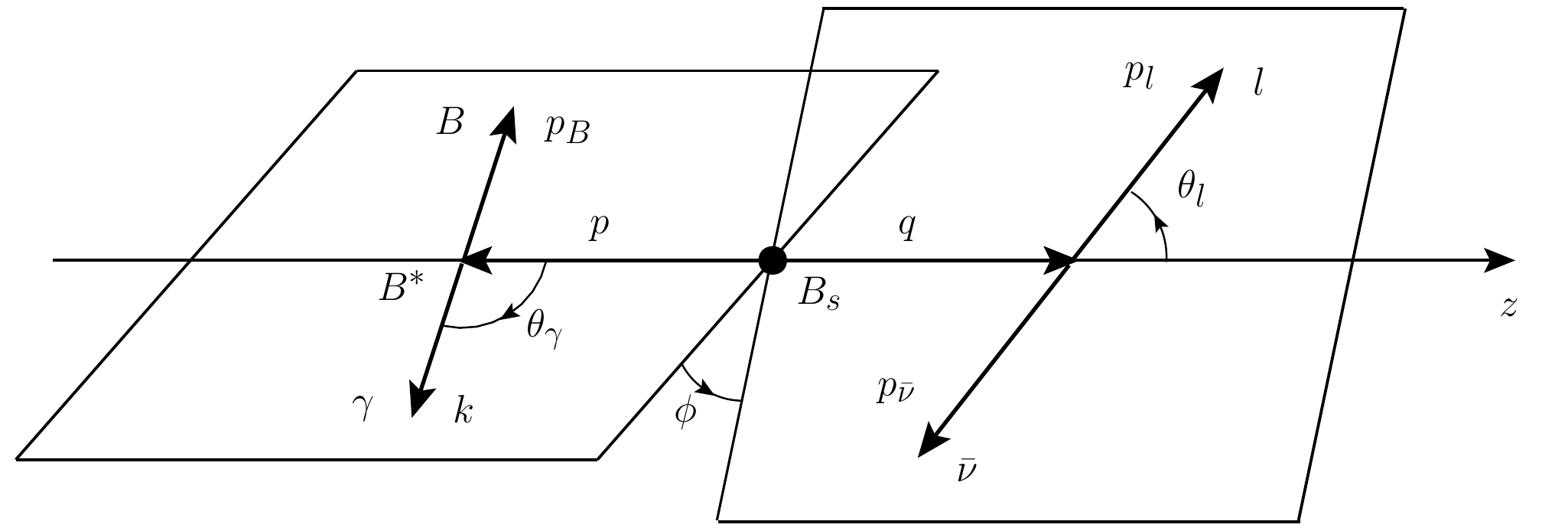}
\caption{The kinematics diagram of $B_{s}\to B^{*}(\to B \gamma)l\bar\nu$, where $p=p_B+p_{\gamma}$ and $q=p_{l}+p_{\bar\nu}$. $\theta_{\gamma}$ is the angular between the moving direction of the photon and the z-axis. $\phi$ is azimuth angle between the lepton pair plane and the $B\gamma$ plane. }
\label{fig:angularDistr}
\end{figure*}
We further consider the four-body decay process: $B_{s}\to B^{*}(\to B \gamma)l\bar\nu$, where the final $B^*$ meson undergoes a radiative decay. The corresponding kinematics diagram is shown in Fig.\ref{fig:angularDistr}, with $p=p_B+p_{\gamma}$ and $q=p_{l}+p_{\bar\nu}$. The angular between the moving direction of the photon and the z-axis is denoted as $\theta_{\gamma}$.
The helicity amplitude of $B_{s}\to B^{*}(\to B \gamma)l\bar\nu$ is expressed as
\begin{align}
{\cal A}_{\rm 4bdy}^{h_{\gamma},h_l}=\frac{G_F}{\sqrt{2}}V_{us}\Big[&H^{\mu}(h_{\gamma})\varepsilon_{\mu}^{V*}(t)B^{h_l}_t\nonumber\\
&-\sum_{\lambda=\pm,0}H^{\mu}(h_{\gamma})\varepsilon_{\mu}^{V*}(\lambda)B^{h_l}_{\lambda}\Big],
\end{align}
where $H_{\mu}(h_{\gamma})=\langle B,\gamma(h_{\gamma})|\bar{u}\gamma_{\mu}(1-\gamma_5)s|B_{s}\rangle$.
The $B^*$ meson  plays the role of a resonant in the $B_{s}\to B^{*}(\to B \gamma)l\bar\nu$ decay, and its contribution can be included by using the Breit-Wigner distribution:
\begin{align}
H^{\mu}(h_{\gamma})\varepsilon_{\mu}^{V*}(h_V)=&\sum_{h_R=\pm,0}\int\frac{d^4 l}{(2\pi)^4}\frac{i}{l^2-m_{B^{*}}^2+im_{B^{*}}\Gamma_{B^{*}}}\nonumber\\
&\times\langle B,\gamma(h_{\gamma})|B^*(l,h_R)\rangle A^{h_R}_{h_V}.
\end{align}
$\Gamma_{B^*}$ denotes the decay width of $B^*$ meson, which is totally dominated by its electromagnetic decay mode $B^*\to B\gamma$. Generally, the decay amplitude of $B^*\to B \gamma$ can be parameterized as
\begin{align}
&\langle B(p_B),\gamma(k,h_{\gamma})|B^*(p,h_R)\rangle\nonumber\\
=&\ (2\pi)^4 \delta^4(p-p_B-k)D^{h_{\gamma}}_{h_R},
\end{align}
with
\begin{align}
D^{h_{\gamma}}_{h_R}=\frac{g_0}{m_{B^*}} \varepsilon^{\mu\nu\rho\sigma}\varepsilon^{\gamma *}_{\mu}(k,h_{\gamma}) p_{\nu}k_{\rho}\varepsilon^{R}_{\sigma}(p,h_R),
\end{align}
where $\varepsilon^{\gamma}$ and $\varepsilon^{R}$ are polarization vectors of the photon and $B^*$ meson, respectively. $g_0$ is a dimensionless coupling constant. The expressions of $D^{h_{\gamma}}_{h_R}$ can be found in the Appendix.  The helicity amplitude becomes
\begin{align}
{\cal A}_{\rm 4bdy}^{h_{\gamma},h_l}=&\frac{G_F}{\sqrt{2}}V_{us}\frac{i}{p^2-m_{B^{*}}^2+im_{B^{*}}\Gamma_{B^{*}}}\nonumber\\
&\times\left[D^{h_{\gamma}}_0 A^0_t B^{h_l}_t-\sum_{\lambda=\pm,0}D^{h_{\gamma}}_{\lambda}A^{\lambda}_{\lambda}B^{h_l}_{\lambda}\right].
\end{align}

All the information about the angular $\theta_{\gamma}$ and $\theta_l$ are contained in the $D^i_j$ and $B^i_j$ functions. The differential decay width of $B_{s}\to B^{*}(\to B \gamma)l\bar\nu$ reads as
\begin{align}
&\frac{d \Gamma_{B_{s}\to B \gamma l\bar\nu}}{d p^2 d q^2 d {\rm cos} \theta_l d {\rm cos} \theta_{\gamma} d\phi}\nonumber\\
=\ &{\cal K}\Big[I_1+I_2 \ {\rm cos}2\theta_l+I_3\  {\rm sin}^2\theta_l{\rm cos}2\phi\nonumber\\
&+I_4\  {\rm sin}2\theta_l{\rm cos}\phi+I_5\  {\rm sin}\theta_l{\rm cos}\phi+I_6\  {\rm cos}\theta_l\nonumber\\
&+I_7\  {\rm sin}\theta_l{\rm sin}\phi+I_8\  {\rm sin}2\theta_l{\rm sin}\phi\nonumber\\
&+I_9\  {\rm sin}^2\theta_l{\rm sin}2\phi\Big],\label{eq:4bdyAngular}
\end{align}
with 
\begin{align}
{\cal K}=\frac{45\  G_F^2 |V_{us}|^2\beta_l^2}{44(4\pi)^5m_{B_s}^3}\frac{\Gamma_{B^*\to B\gamma}\ q^2 \sqrt{{\tilde\lambda}_{B_s}p^2}}{(p^2-m_{B^*}^2)^2+m_{B^*}^2\Gamma_{B^*\to B\gamma}^2}
\end{align}
where ${\tilde\lambda}_{B_s}=[m_{B_s}^2-(\sqrt{p^2}+\sqrt{q^2})^2][m_{B_s}^2-(\sqrt{p^2}-\sqrt{q^2})^2]$. $\Gamma_{B^*\to B\gamma}$ is the decay width of $B^*$ which is dominated by the electromagnetic decay mode $B^*\to B\gamma$. The $I_i$ s are functions of $p^2, q^2$ and $\theta_{\gamma}$ :
\begin{align}
I_1&=-(1+{\hat m}_l^2){\rm cos}2\theta_{\gamma}|A^0_0|^2-2{\hat m}_l^2{\rm cos}2\theta_{\gamma}|A^0_t|^2\nonumber\\
I_2&=\beta_l{\rm cos}2\theta_{\gamma}|A^0_0|^2\nonumber\\
&+\frac{1}{4}\beta_l(1-6\ {\rm cos}^2\theta_{\gamma}+8\ {\rm cos}^4\theta_{\gamma})(|A_{\parallel}|^2+|A_{\perp}|^2)\nonumber\\
I_3&=\frac{1}{2}\beta_l(1-10\ {\rm cos}^2\theta_{\gamma}+8\ {\rm cos}^4\theta_{\gamma})(|A_{\parallel}|^2-|A_{\perp}|^2)\nonumber\\
I_4&=\frac{3}{\sqrt 2}\beta_l {\rm sin}2\theta_{\gamma}{\rm Re}(A^0_0 A_{\parallel}^*)\nonumber\\
I_5&=-3\sqrt 2 {\rm sin}2\theta_{\gamma}\left[{\rm Re}(A^0_0 A_{\perp}^*)-{\hat m}_l^2{\rm Re}(A^0_t A_{\parallel}^*)\right]\nonumber\\
I_6&=4 {\hat m}_l^2 {\rm cos}2\theta_{\gamma}{\rm Re}(A^0_0 A^{0*}_t)\nonumber\\
&-2(1-6\ {\rm cos}^2\theta_{\gamma}+8\ {\rm cos}^4\theta_{\gamma}){\rm Re}(A_{\parallel} A_{\perp}^*)\nonumber\\
I_7&=-3\sqrt 2 {\rm sin}2\theta_{\gamma}\left[{\rm Im}(A^0_0 A_{\parallel}^*)-{\hat m}_l^2{\rm Im}(A^0_t A_{\perp}^*)\right]\nonumber\\
I_8&=\frac{3}{\sqrt 2}\beta_l{\rm sin}2\theta_{\gamma}{\rm Im}(A^0_0 A_{\perp}^*)\nonumber\\
I_9&=\beta_l(1-10\ {\rm cos}^2\theta_{\gamma}+8\ {\rm cos}^4\theta_{\gamma}){\rm Im}(A_{\parallel} A_{\perp}^*).\label{eq:4bdyAngularIs}
\end{align}

To investigate the angular distribution in terms of $\theta_l$, one has to integrate out the $\theta_{\gamma}$ and $\phi$, which leads to
\begin{align}
&\frac{d \Gamma_{B_{s}\to B \gamma l\bar\nu}}{d p^2 d q^2 d {\rm cos} \theta_l}\nonumber\\
=\ &2\pi {\cal K}\int d{\cos}\theta_{\gamma}[I_1+I_2 {\cos}2\theta_l+I_6 {\cos}\theta_l].
\end{align}
The normalized FB asymmetry in terms of $\theta_l$ with $p^2$ and ${\cos}\theta_{\gamma}$ being integrated out is defined as
\begin{align}
A^{4,l}_{FB}=&\frac{\left(\int^1_0-\int^0_{-1}\right)d {\rm cos} \theta_l \frac{d \Gamma_{B_{s}\to B \gamma l\bar\nu}}{d q^2 d {\rm cos} \theta_l}}{\int^1_{-1} d {\rm cos} \theta_l \frac{d \Gamma_{B_{s}\to B \gamma l\bar\nu}}{d q^2 d {\rm cos} \theta_l}}\nonumber\\
=&\frac{3}{2}\frac{\int d p^2 d{\cos}\theta_{\gamma}\ {\cal K}I_6}{\int d p^2 d{\cos}\theta_{\gamma}\ {\cal K} (3 I_1 -I_2)},\label{eq:FBasymBstoBgamma}
\end{align}
which is a function of $q^2$. Similarly, integrating out the $\theta_{l}$ and $\phi$ one can obtain the distribution of $\theta_{\gamma}$:
\begin{align}
&\frac{d \Gamma_{B_{s}\to B \gamma l\bar\nu}}{d p^2 d q^2 d{\cos}\theta_{\gamma}}= \frac{4}{3}\pi {\cal K}\left(3 I_1-I_2\right).
\end{align}
It can be found from Eq.~(\ref{eq:4bdyAngularIs}) that the $I_1, I_2$ are even functions of ${\cos}\theta_{\gamma}$, so that the $\theta_{\gamma}$ distribution has no FB asymmetry. Besides $\theta_{\gamma}$ and $\theta_{l}$, one can also investigate the distribution of $\phi$, which reads as
\begin{align}
\frac{d \Gamma_{B_{s}\to B \gamma l\bar\nu}}{d q^2 d\phi}=&\  a_{\phi}+b_{\phi}^{c} \cos \phi+b_{\phi}^{s} \sin \phi\nonumber\\
&+c_{\phi}^{c} \cos 2 \phi+c_{\phi}^{s} \sin 2 \phi,
\end{align}
where the coefficients above are
\begin{align}
a_{\phi}&=\int dp^2  d{\cos}\theta_{\gamma}\ {\cal K} \left(I_1-\frac{2}{3}I_2\right),\nonumber\\
b_{\phi}^{c}&=\frac{\pi}{2}\int dp^2 d{\cos}\theta_{\gamma}\ {\cal K}\   I_5,\nonumber\\
b_{\phi}^{s}&=\frac{\pi}{2}\int dp^2  d{\cos}\theta_{\gamma}\ {\cal K}\   I_7,\nonumber\\
c_{\phi}^{c}&=\frac{4}{3}\int dp^2  d{\cos}\theta_{\gamma}\ {\cal K}\   I_3,\nonumber\\
c_{\phi}^{s}&=\frac{4}{3}\int dp^2 d{\cos}\theta_{\gamma}\ {\cal K}\   I_9.\label{eq:phicoeffs}
\end{align}
Note that $b_{\phi}^{s}= c_{\phi}^{s}=0$ because the $A^i_j$ s are purely imaginary values so that the $I_7, I_9$ vanish. $a_{\phi}$ and $c_{\phi}^{c}$ are even functions of ${\cos}\theta_{\gamma}$ so that have no FB asymmetry. $b_{\phi}^{c}$ is an odd function of ${\cos}\theta_{\gamma}$ and we can define the corresponding  FB asymmetry as
\begin{align}
&A_{FB}^{b^c}=\frac{\left(\int^1_0-\int^0_{-1}\right)d {\rm cos} \theta_{\gamma} \ b_{\phi}^{c}}{\frac{d}{dq^2}{\Gamma_{B_{s}\to B \gamma l\bar\nu}}}\nonumber\\
=\ &\frac{3}{8}\frac{\int dp^2 d{\cos}\theta_{\gamma}\ {\cal K}\  I_5}{\int d p^2 d{\cos}\theta_{\gamma}\ {\cal K}\  (3 I_1 -I_2)}.
\end{align}
Note that since the integration of  $b_{\phi}^{c}$  in the region $-1<{\cos}\theta_{\gamma}<1$ is zero, the normalization factor in the denominator is chosen as  $(d/dq^2) \Gamma_{B_{s}\to B \gamma l\bar\nu}$ instead of $\int^1_{-1}d {\rm cos} \theta_{\gamma} \ b_{\phi}^{c}$. One can do the same normalization for $a_{\phi}$ and $c^c_{\phi}$, which reads as
\begin{align}
a_{\phi}\to \frac{a_{\phi}}{\frac{d}{dq^2}{\Gamma_{B_{s}\to B \gamma l\bar\nu}}},\ \ c^c_{\phi}\to \frac{c^c_{\phi}}{\frac{d}{dq^2}{\Gamma_{B_{s}\to B \gamma l\bar\nu}}}.
\end{align}

\section{Numerical Results}
\subsection{Transition Form Factors}
In the covariant LFQM, we use the constituent quark masses (in units of GeV): $m_{u}=m_{d}=0.25,\ m_{s}=0.37,\ m_{b}=4.8$, 
which were widely used in various of heavy meson decays~\cite{Lu:2007sg,Wang:2007sxa,Wang:2008xt,Wang:2008ci,Wang:2009mi,Chen:2009qk,Li:2010bb,Verma:2011yw}. The masses of the $B_s$ and $B^{(*)}$  are taken as (in units of GeV): $m_{B_{s}}=5.367,\ m_B=5.279,\ m_{B^*}=5.325$~\cite{Agashe:2014kda}. Note that since the phase spaces of $B_s\to B^{(*)}$ and $D_s\to D$  are too small so that only electron can be produced in such processes. Therefore we simply set $m_l=m_e=0$ in the numerical calculation.  The shape parameters for the $B_s,B^{(*)}$ mesons are taken as $\beta_{B_s}=0.623$ GeV, $\beta_{B}=\beta_{B^*}=0.623$ GeV \cite{Shi:2016gqt}, which are fitted from the $B_s, B^{(*)}$ meson decay constants. The shape parameters for the $D_s,D$ mesons are taken as $\beta_{D_s}=0.537$ GeV and $\beta_{D}=0.464$ GeV \cite{Chang:2018zjq}.

\begin{table}
  \caption{Transition form factors for $B_s\to B^{(*)}$ and $D_s\to D$ at $q^2=0$ in the two parameterization schemes.}
\label{Tab:formfactors}
\begin{tabular}{|c|cccccc|}
\hline 
\hline
Scheme 1 & $F_0(0)$ & $F_1(0)$ & $V(0)$ & $A_0(0)$ & $A_1(0)$ & $A_2(0)$ \tabularnewline
\hline 
$B_s\to B^{(*)}$ & $0.985$ & $0.985$ & $8.912$ & $0.595$ & $0.593$ & $0.205$ \tabularnewline
$D_s\to D$ & $0.98$ & $0.98$ & &  &  &  \tabularnewline
\hline 
Scheme 2 & $f_{+}(0)$ & $f_{-}(0)$ & $g(0)$ & $f(0)$ & $a_{+}(0)$ & $a_{-}(0)$ \tabularnewline
\hline 
$B_s\to B^{(*)}$ & $0.985$ & $-0.879$ & $-0.834$ & $-6.346$ & $0.019$ & $-9.724$ \tabularnewline
$D_s\to D$ & $0.98$ & $-0.19$ &  &  &  &  \tabularnewline
\hline 
\end{tabular}
\end{table}
\begin{table}
  \caption{Reduced $B_s\to B^{(*)}$ transition form factors from heavy quark symmetry.}
\label{Tab:HQETformfactors}
\begin{tabular}{|c|cccccc|}
\hline 
\hline
Scheme 2 & $f_{+}(0)$ & $f_{-}(0)$ & $g(0)$ & $f(0)$ & $a_{+}(0)$ & $a_{-}(0)$ \tabularnewline
\hline 
HQS & $1.0$ & $0$ & $0$ & $-10.6$ & $0$ & $0$ \tabularnewline
\hline 
\end{tabular}
\end{table}
Since the phase space of $q^2$ is tiny in the $s\to u$ transition, the form factors defined in Eq.~(\ref{eq:formfactors}) varies little in the physical $q^2$ regions. Therefore, we can use single pole model to parameterize the form factors, which reads as:
\begin{align}
F_V(q^2)=\frac{F_V(0)}{1-\frac{q^2}{m_{K^*}^2}},\ \ F_A(q^2)=\frac{F_A(0)}{1-\frac{q^2}{m_{K}^2}},
\end{align}
where $F_V=F_{0,1}, V$ denote the form factors induced by the vector current, while $F_A=A_{0,1,2}$ denote the form factors induced by the axial-vector current. The numerical results for the form factors are listed in Table \ref{Tab:formfactors}, where Scheme 1,2 denote two parameterization given by Eq.~(\ref{eq:formfactors}) and Eq.~(\ref{eq:formfactors2}).  For completeness we have also calculated the form factors for $D_s\to D$ transition. The $D_s\to D^*$ transition is forbidden because $m_{D_s}<m_{D^*}$. 

Now using the results listed in Table \ref{Tab:formfactors},  we can test the heavy quark symmetry (HQS) in the heavy flavor conserved processes. According to Ref. \cite{Faller:2015oma}, in the $v\to v^{\prime}$ limit, the leading contribution to matrix elements given in Eq.~(\ref{eq:formfactors}) reads as:
\begin{align}
\frac{\langle B(P^{\prime})|\bar{u}\gamma_{\mu}s|B_{s}(P)\rangle}{\sqrt{m_{B_s}m_B}} &=(v+v^{\prime})_{\mu}\Phi_+(w),\nonumber \\
\frac{\langle B^*(P^{\prime})|\bar{u}\gamma_{\mu}\gamma_5 s|B_{s}(P)\rangle}{\sqrt{m_{B_s}m_B}} &= i(1+w)\varepsilon^*_{\mu}(v^{\prime})\Phi_{A_1}(w),
\label{eq:HQETformfactors}
\end{align}
where $P=m_{B_s}v$, $P^{\prime}=m_{B^{(*)}}v^{\prime}$ and $w=v\cdot v^{\prime}$. The vector current conservation leads to $\Phi_+(1)=1$, while the the spin counting estimated by Ref. \cite{Faller:2015oma} leads to $\Phi_{A_1}(1)=1$. Comparing Eq.~(\ref{eq:formfactors}) with Eq.~(\ref{eq:HQETformfactors}), and ignoring the slight $q^2$ dependence, one can deduce that $f_+(0)=1$, $f(0)=-2m_{B_s}$ and  $f_{-}(0)=g(0)=a_{+}(0)=a_-(0)=0$, which are listed in Table \ref{Tab:HQETformfactors}. It can be found that the $f_{+}(0)$ in Table \ref{Tab:formfactors} perfectly agrees with the HQS prediction, and the $g(0), f(0), a_+(0)$ are also consistent with HQS. However, $f_{-}(0)$ and $a_{-}(0)$  are much larger than the HQS prediction value. The reason is that both $f_{-}$ and $a_{-}$ are coefficients of $q_{\mu}=(P-P^{\prime})_{\mu}$, which vanishes in the $v\to v^{\prime}$ limit so that  the $f_{-}, a_{-}$ terms are absent in the HQS analysis.

\begin{figure*}[htp]
\includegraphics[scale=0.38]{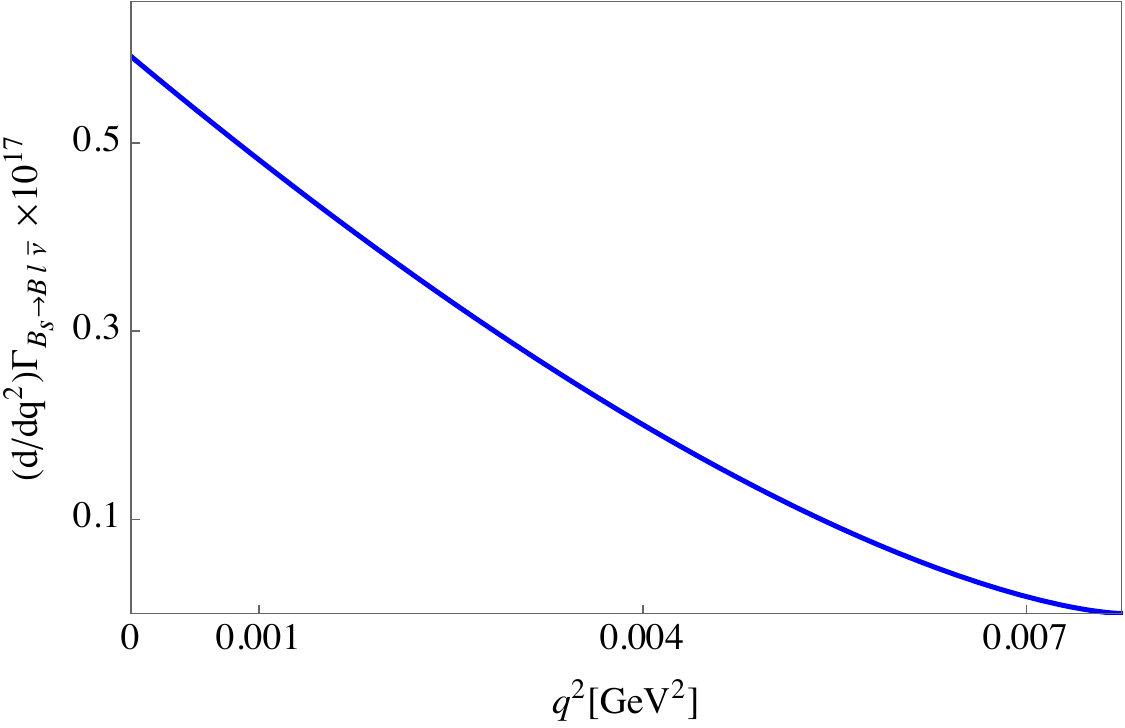}
\includegraphics[scale=0.38]{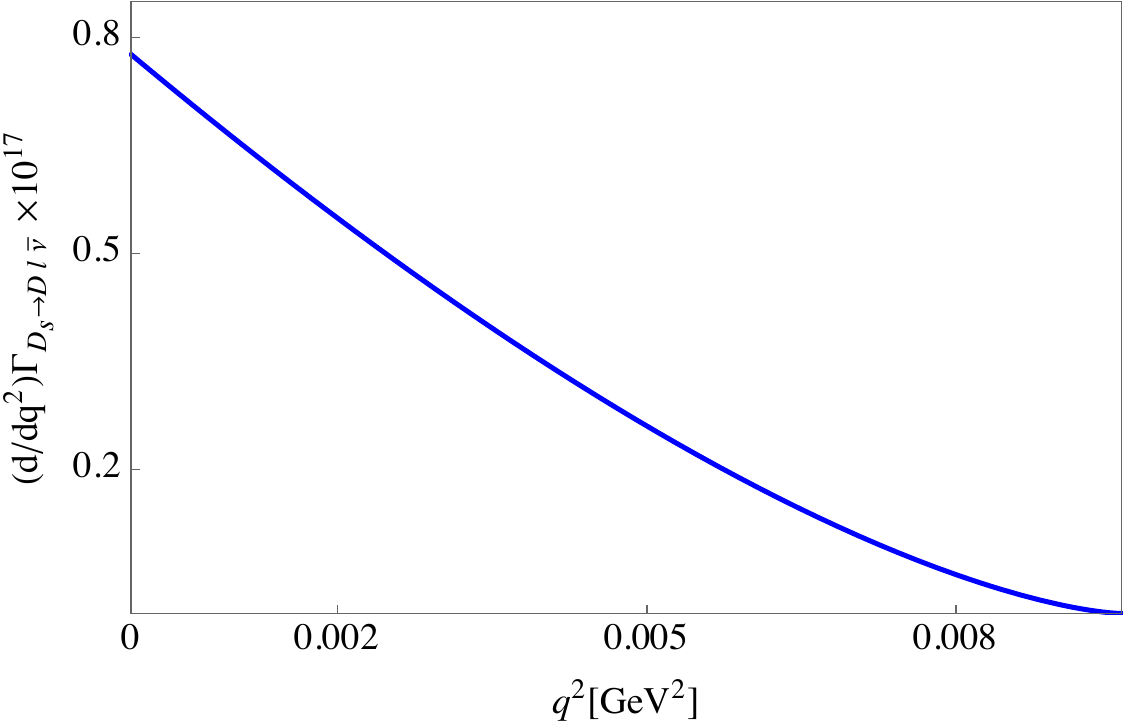}
\caption{The differential decay widths of $B_s\to B l \bar\nu$ (left) and $D_s\to Dl \bar\nu$ (right) as functions of $q^2$.}
\label{fig:Width3bodyVsq2}
\end{figure*}
\begin{figure*}[htp]
\includegraphics[scale=0.38]{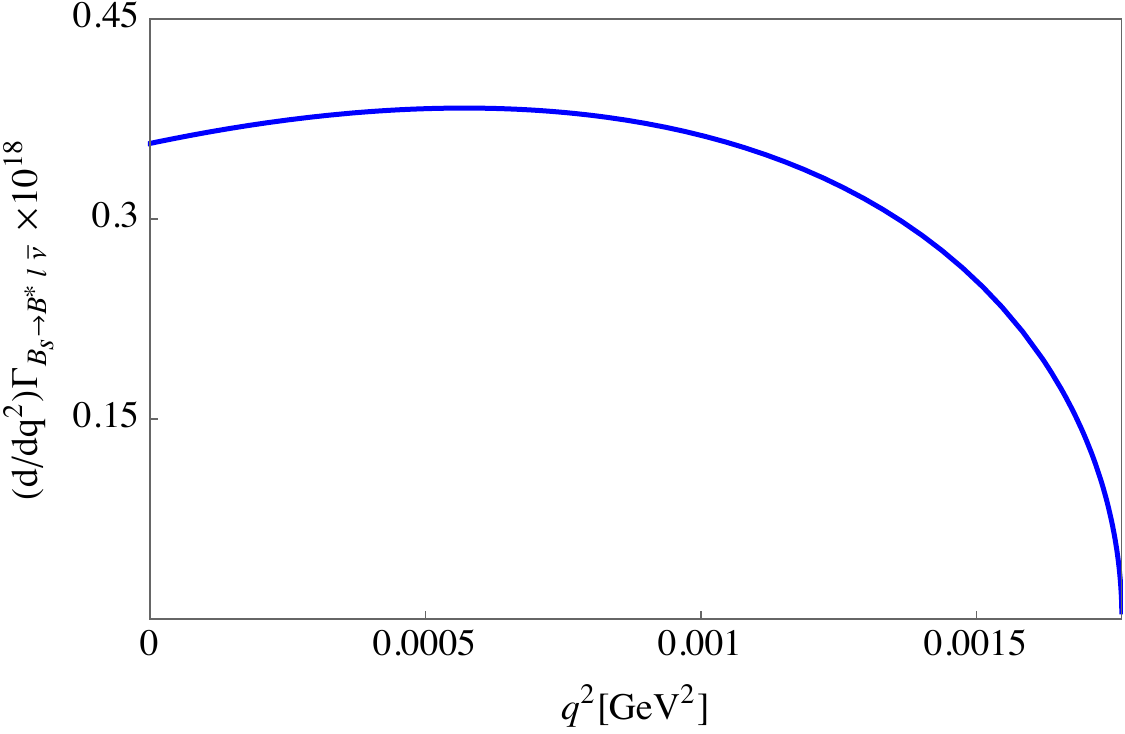}
\includegraphics[scale=0.39]{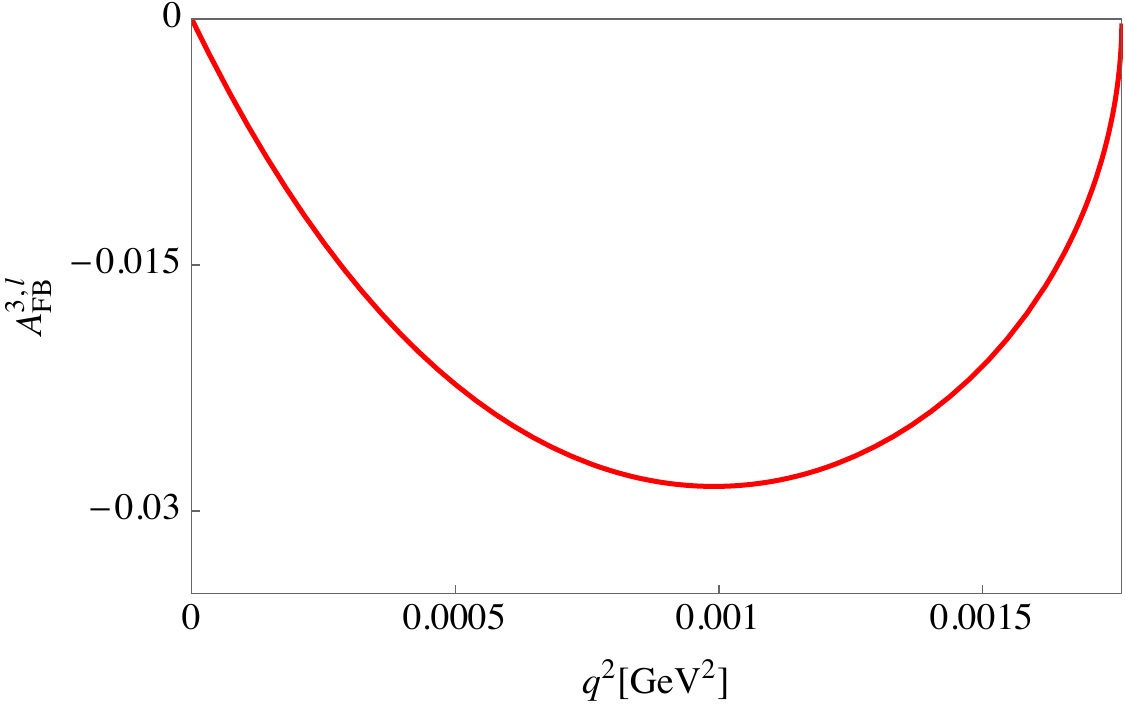}
\caption{The differential decay widths of $B_s\to B^* l \bar\nu$ (left) and the corresponding normalized FB asymmetry (right) as functions of $q^2$.}
\label{fig:AFB3bodyVsq2}
\end{figure*}

\subsection{Branching ratios and angular Distributions}
The differential decay widths of $B_s\to B^{(*)}l \bar\nu$ and $D_s\to Dl \bar\nu$ as functions of $q^2$ are shown in Fig.~\ref{fig:Width3bodyVsq2}. According to Eq.~(\ref{eq:FBasymBstoBstar}), the normalized FB asymmetry of  $B_s\to B^{*}l \bar\nu$ as a function of $q^2$ is shown in Fig.~\ref{fig:AFB3bodyVsq2} shows the differential decay widths of $B_s\to B^{*}l \bar\nu$ and the corresponding normalized FB asymmetry as functions of $q^2$, respectively. The total branching fractions of $B_s\to B^{(*)}l \bar\nu$ and $D_s\to Dl \bar\nu$ are obtained as
\begin{align}
{\cal B}(B_s\to B l \bar\nu)&=4.26\times 10^{-8},\nonumber\\
{\cal B}(D_s\to D l \bar\nu)&=2.30\times 10^{-8},\nonumber\\
{\cal B}(B_s\to B^* l \bar\nu)&=1.34\times 10^{-9}.
\end{align}
Using the measured cross section of $B_s$ production: $(0.340\pm0.016){\rm nb}$ \cite{BaBar:2014omp},  the total number of $B_s$ is $4.13\times 10^7$ with $121.4{\rm fb}^{-1}$ $\Upsilon(5S)$ data in Belle. Then the number of  $B_s\to B l\bar\nu$ events is estimated to be $1.76$.  Though it is still not easy to be measured in the Belle experiment,  the branching fractions of $B_s\to B^{(*)}l\bar\nu$ are possible to be tested in the future with the increasing amount of  $B_s$ samples from the updating  Belle experiment.

\begin{figure*}[htp]
\includegraphics[scale=0.385]{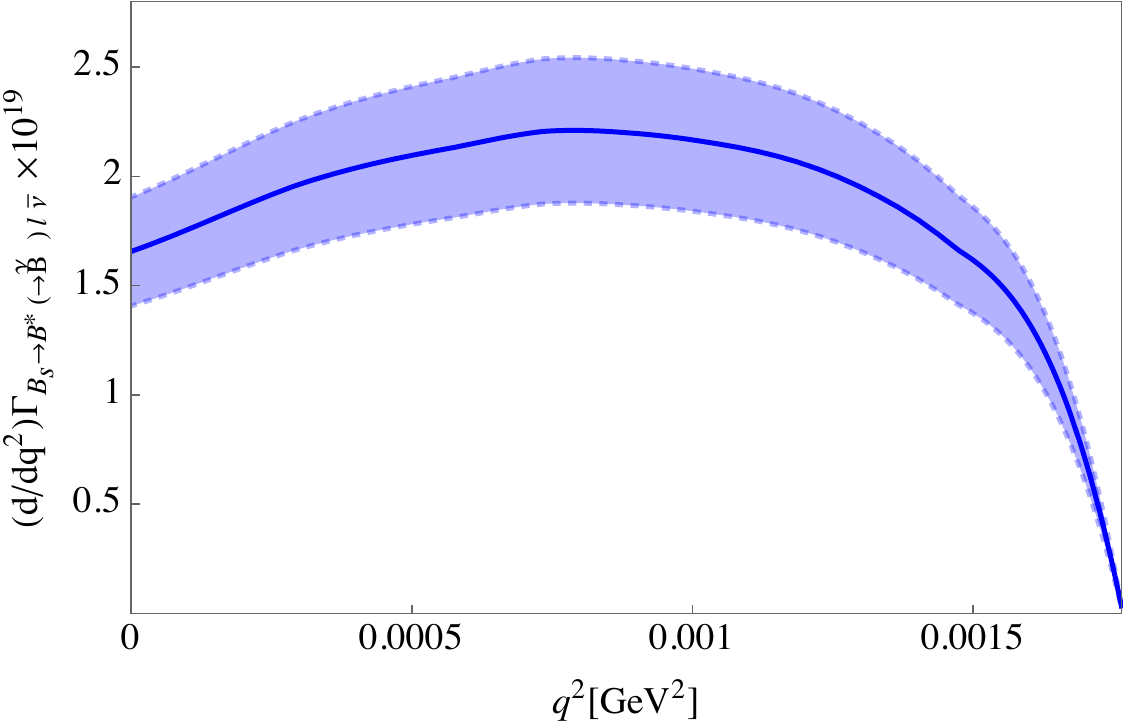}
\includegraphics[scale=0.4]{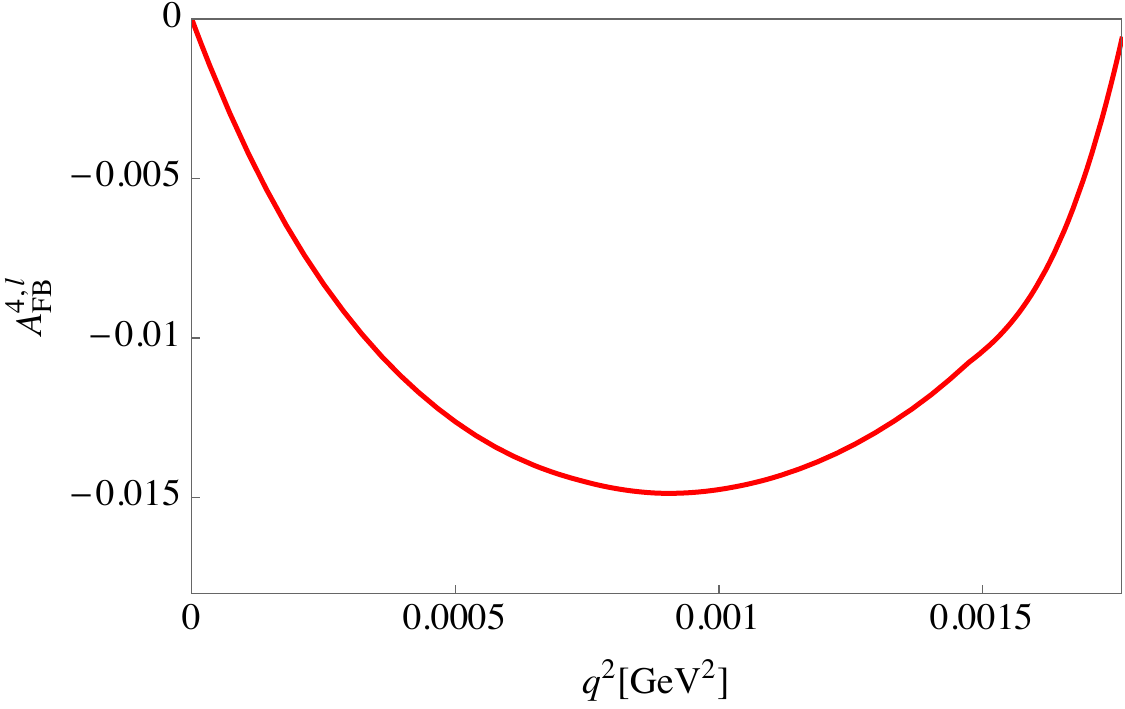}
\caption{The differential decay widths of $B_{s}\to B^{*}(\to B \gamma)l\bar\nu$ (left) and corresponding normalized FB asymmetry (right) as functions of $q^2$.}
\label{fig:AFB4bodyVsq2}
\end{figure*}
The differential decay widths of $B_{s}\to B^{*}(\to B \gamma)l\bar\nu$ and corresponding normalized FB asymmetry $A^{4,l}_{FB}$ defined in Eq.~(\ref{eq:FBasymBstoBgamma}) are shown in Fig.~\ref{fig:AFB4bodyVsq2}, where the decay width of $B^*$ is taken as $\Gamma_{B^*\to B\gamma}=372\pm 56$ eV \cite{Ivanov:2022nnq}. The error band in the left panel of Fig.~\ref{fig:AFB4bodyVsq2} represents the uncertainty from $\Gamma_{B^*\to B\gamma}$. The decay branching fraction of $B_{s}\to B^{*}(\to B \gamma)l\bar\nu$ can be obtained as 
\begin{align}
{\cal B}(B_{s}\to B^{*}(\to B \gamma)l\bar\nu)&=(6.0\pm 0.9)\times 10^{-11}.
\end{align}

\begin{figure*}[htp]
\includegraphics[scale=0.3]{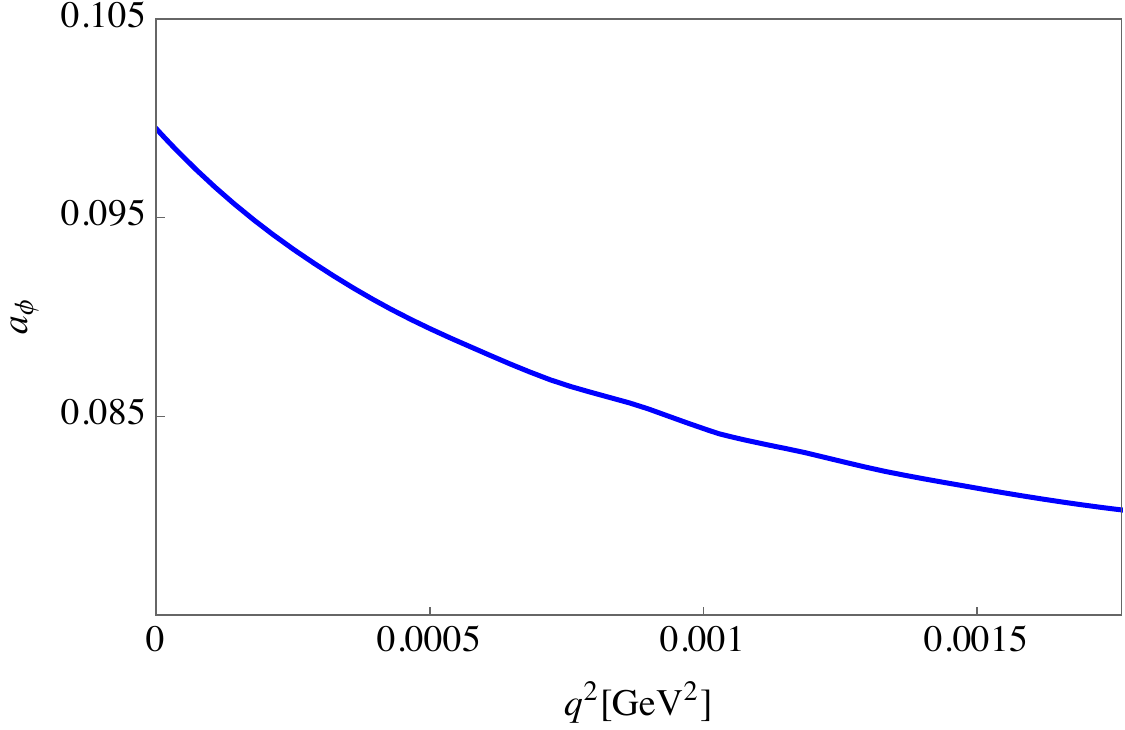}
\includegraphics[scale=0.3]{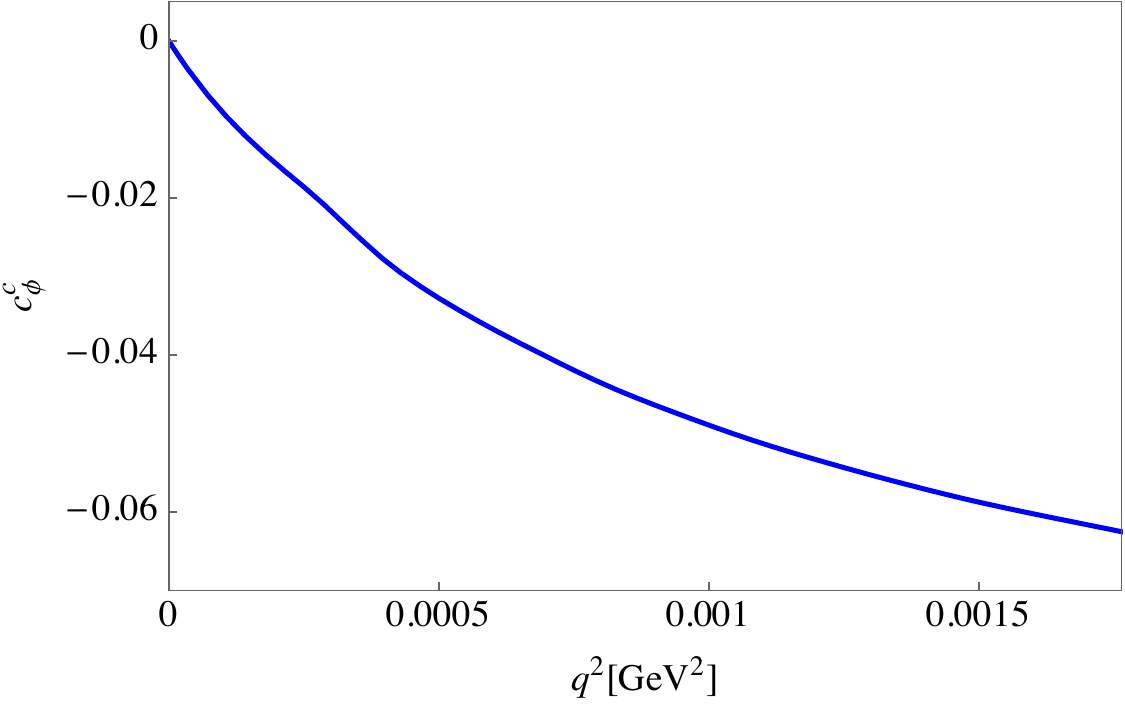}
\includegraphics[scale=0.3]{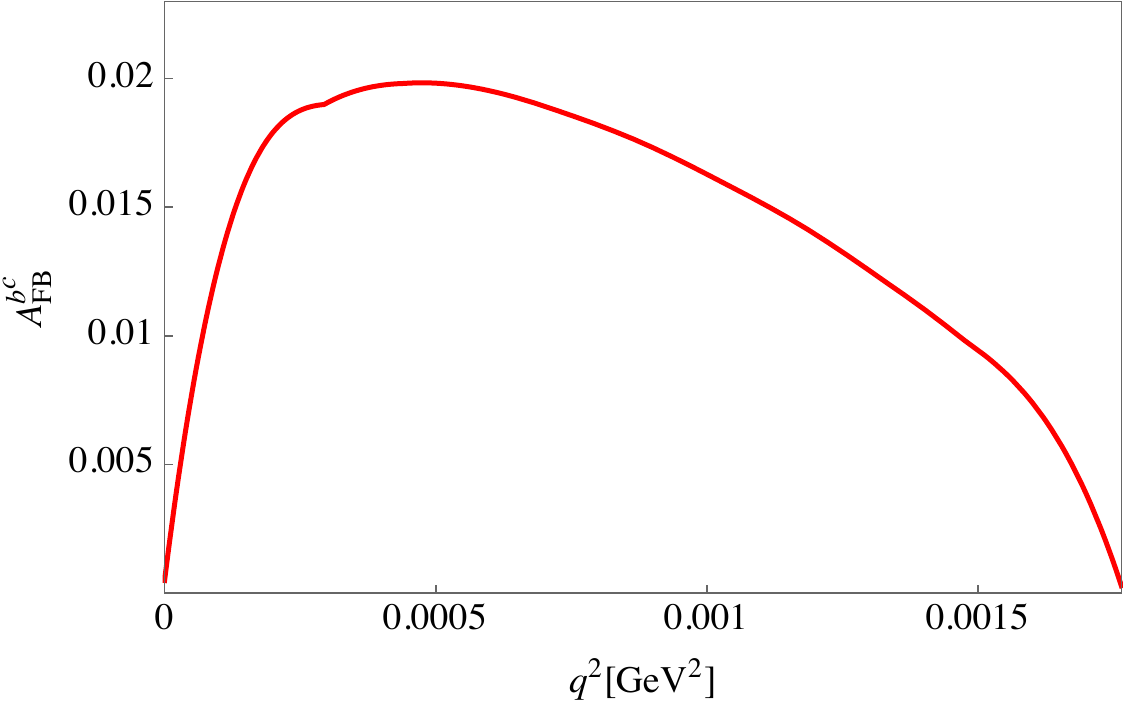}
\caption{The differential decay widths of $B_{s}\to B^{*}(\to B \gamma)l\bar\nu$ (left) and corresponding normalized FB asymmetry (right) as functions of $q^2$.}
\label{fig:phi4bodyVsq2}
\end{figure*}
In terms of FB asymmetry, the $\Gamma_{B^*\to B\gamma}$ in the nominator and denominator of Eq.~(\ref{eq:FBasymBstoBgamma}) cancel with each other, while its uncertainty has little effect on the integration of the Breit-Wigner distribution. Therefore the error band of $A^{4,l}_{FB}$ can be neglected.
The normalized $a_{\phi}, c^c_{\phi}$ and $A_{FB}^{b^c}$ as functions of $q^2$ are shown in Fig.~\ref{fig:phi4bodyVsq2}.

\section*{Conclusion}
In this work, we studied the heavy flavor conserved semi-leptonic decay $B_s\to B\ell\nu$ in the covariant light front approach.  We used covariant light front quark model to calculate the transition form factors of $B_s\to B^{(*)}$ as well as $D_s\to D$, and the results are consistent with the leading power predictions from HQS.  We performed an angular distribution analysis  on the $B_s\to B^{(*)}l\bar\nu$ decay by investigating the forward-backward asymmetry of the lepton. The angular distribution of $B_{s}\to B^{*}(\to B \gamma)l\bar\nu$ decay is studied both through the lepton forward-backward asymmetry and the azimuth angle.  The obtained branching fractions of $B_s\to Bl\bar\nu$ and $B_s\to B^{*}l\bar\nu$ are  at the order $10^{-8}$ and $10^{-9}$, respectively. The obtained branching fraction of $B_{s}\to B^{*}(\to B \gamma)l\bar\nu$ is at the order $10^{-11}$.  Using the measured cross section of $B_s$ production, we deduced that the number of  $B_s\to B l\bar\nu$ events is $1.76$, which is possible to be tested in the future with the increasing amount of  $B_s$ samples produced by the updating  Belle experiment.

\section*{Acknowledgements}
We thank Prof. W. Wang in Shanghai Jiao Tong University, Prof. C.P. Shen in Fudan University and Prof. G. Li in Sun Yat-sen University for very useful discussion. 
The work of Y.J. Shi is supported by Opening Foundation of Shanghai Key Laboratory of Particle Physics and Cosmology under Grant No.22DZ2229013-2. 
The work of Z.P. Xing is supported by China Postdoctoral Science Foundation under Grant No.2022M72210.

\begin{appendix}

\section{Expression of form factors}\label{ExpressFF}
Here we present the analytical expressions of the transition form factors for $B_s\to B^{(*)}$, which were derived in Refs.~\cite{Jaus:1999zv,Cheng:2003sm}. For convenience we use an alternative parameterization scheme for the form factors:
\begin{widetext}
\begin{align}
&\langle B(P^{\prime})|\bar{u}\gamma_{\mu}s|B_{s}(P)\rangle = {\bar P}_{\mu} f_{+}(q^2)+q_{\mu} f_{-}(q^2),\nonumber \\
&\langle B^*(P^{\prime},\varepsilon^{\prime})|\bar{u}\gamma_{\mu}s|B_{s}(P)\rangle=  \epsilon_{\mu\nu\alpha\beta}\varepsilon^{\prime}{}^{*\nu}{\bar P}^{\alpha}q^{\beta} g(q^2),\nonumber \\
&\langle B^*(P^{\prime},\varepsilon^{\prime})|\bar{u}\gamma_{\mu}\gamma_{5} s|B_{s}(P)\rangle=-i\left\{\varepsilon^{\prime *}_{\mu}f(q^2)+\varepsilon^{\prime *}\cdot {\bar P}\left[{\bar P}_{\mu} a_{+}(q^2)+q_{\mu}a_{-}(q^2)\right]\right\},\label{eq:formfactors2}
\end{align}
\end{widetext}
where ${\bar P}=P+P^{\prime}$. The relation between the form factors defined in Eq.~(\ref{eq:formfactors}) and those in Eq.~(\ref{eq:formfactors2}) as:
\begin{align}
F_{1}\left(q^{2}\right)=& f_{+}\left(q^{2}\right), \nonumber\\
F_{0}\left(q^{2}\right)=& f_{+}\left(q^{2}\right)+\frac{q^{2}}{q \cdot P} f_{-}\left(q^{2}\right), \nonumber\\
V\left(q^{2}\right)=&-\left(m_{B_{s}}+m_{B^*}\right) g\left(q^{2}\right), \nonumber\\
A_{1}\left(q^{2}\right)=&-\frac{f\left(q^{2}\right)}{m_{B_{s}}+m_{B^*}}, \nonumber\\
A_{2}\left(q^{2}\right)=&\left(m_{B_{s}}+m_{B^*}\right) a_{+}\left(q^{2}\right), \nonumber\\
A_{0}\left(q^{2}\right)=&\frac{m_{B_{s}}+m_{B^*}}{2 m_{B^*}} A_{1}\left(q^{2}\right)-\frac{m_{B_{s}}-m_{B^*}}{2 m_{B^*}} A_{2}\left(q^{2}\right)\nonumber\\
&-\frac{q^{2}}{2 m_{B^*}} a_{-}\left(q^{2}\right).
\end{align}
The analytical expressions of the transition form factors defined above read as
\begin{widetext}
\begin{eqnarray}
f_{+}(q^{2}) & = & \frac{N_{c}}{16\pi^{3}}\int dx_{2}d^{2}{\vec k}_{\perp}\frac{h_{{\cal M}}h_{{\cal M}^{\prime}}}{x_{2}\hat{N}_{1}^{\prime}\hat{N}_{1}^{\prime\prime}}\Bigg[x_{1}(M_{0}^{2}+M_{0}^{\prime2})+x_{2}q^{2}-x_{2}(m_{1}-m_{1}^{\prime})^{2}-x_{1}(m_{1}-m_{2})^{2}-x_{1}(m_{1}^{\prime}-m_{2})^{2}\Bigg],\nonumber \\
f_{-}(q^{2}) & = & \frac{N_{c}}{16\pi^{3}}\int dx_{2}d^{2}{\vec k}_{\perp}\frac{2h_{{\cal M}}h_{{\cal M}^{\prime}}}{x_{2}\hat{N}_{1}^{\prime}\hat{N}_{1}^{\prime\prime}}\Bigg\{-x_{1}x_{2}M^{2}-p_{\perp}^{\prime2}-m_{1}m_{2}+(m_{1}^{\prime}-m_{2})\nonumber \\
 &  & \times(x_{2}m_{1}+x_{1}m_{2})+2\frac{q\cdot P}{q^{2}}\left(p_{\perp}^{\prime2}+2\frac{({\vec k}_{\perp}\cdot{\vec q}_{\perp})^{2}}{q^{2}}\right)+2\frac{({\vec k}_{\perp}\cdot{\vec q}_{\perp})^{2}}{q^{2}}\nonumber \\
 &  & -\frac{{\vec k}_{\perp}\cdot{\vec q}_{\perp}}{q^{2}}\Bigg[M^{\prime2}-x_{2}(q^{2}+q\cdot P)-(x_{2}-x_{1})M^{2}+2x_{1}M_{0}^{2} -2(m_{1}-m_{2})(m_{1}+m_{1}^{\prime})\Bigg]\Bigg\},\\
 g(q^{2}) & = & -\frac{N_{c}}{16\pi^{3}}\int dx_{2}d^{2}{\vec k}_{\perp}\frac{2h_{{\cal M}}h_{V}^{\prime\prime}}{x_{2}\hat{N}_{1}^{\prime}\hat{N}_{1}^{\prime\prime}}\Bigg\{ x_{2}m_{1}+x_{1}m_{2}+(m_{1}-m_{1}^{\prime})\frac{{\vec k}_{\perp}\cdot{\vec q}_{\perp}}{q^{2}}
 +\frac{2}{w_{V}^{\prime\prime}}\Bigg[p_{\perp}^{\prime2}+\frac{({\vec k}_{\perp}\cdot{\vec q}_{\perp})^{2}}{q^{2}}\Bigg]\Bigg\},\\
 f(q^{2}) & = & \frac{N_{c}}{16\pi^{3}}\int dx_{2}d^{2}{\vec k}_{\perp}\frac{h_{{\cal M}}h_{V}^{\prime\prime}}{x_{2}\hat{N}_{1}^{\prime}\hat{N}_{1}^{\prime\prime}}\Bigg\{2x_{1}(m_{2}-m_{1})(M_{0}^{2}+M_{0}^{\prime2})-4x_{1}m_{1}^{\prime}M_{0}^{2}\nonumber \\
 &  & +2x_{2}m_{1}q\cdot P+2m_{2}q^{2}-2x_{1}m_{2}(M^{2}+M^{\prime2})+2(m_{1}-m_{2})(m_{1}+m_{1}^{\prime})^{2}\nonumber \\
 &  & +8(m_{1}-m_{2})\Bigg[p_{\perp}^{\prime2}+\frac{({\vec k}_{\perp}\cdot{\vec q}_{\perp})^{2}}{q^{2}}\Bigg]+2(m_{1}+m_{1}^{\prime})(q^{2}+q\cdot P)\frac{{\vec k}_{\perp}\cdot{\vec q}_{\perp}}{q^{2}}\nonumber \\
 &  & -4\frac{q^{2}p_{\perp}^{\prime2}+({\vec k}_{\perp}\cdot{\vec q}_{\perp})^{2}}{q^{2}w_{V}^{\prime\prime}}\Bigg[2x_{1}(M^{2}+M_{0}^{2})-q^{2}-q\cdot P-2(q^{2}+q\cdot P)\frac{{\vec k}_{\perp}\cdot{\vec q}_{\perp}}{q^{2}}\nonumber \\
 &  & -2(m_{1}-m_{1}^{\prime})(m_{1}-m_{2})\Bigg]\Bigg\},\\
 a_{+}(q^{2}) & = & \frac{N_{c}}{16\pi^{3}}\int dx_{2}d^{2}{\vec k}_{\perp}\frac{2h_{{\cal M}}h_{V}^{\prime\prime}}{x_{2}\hat{N}_{1}^{\prime}\hat{N}_{1}^{\prime\prime}}\Bigg\{(x_{1}-x_{2})(x_{2}m_{1}+x_{1}m_{2})
 -[2x_{1}m_{2}+m_{1}^{\prime}+(x_{2}-x_{1})m_{1}]\frac{{\vec k}_{\perp}\cdot{\vec q}_{\perp}}{q^{2}}\nonumber \\
 &  & -2\frac{x_{2}q^{2}+{\vec k}_{\perp}\cdot{\vec q}_{\perp}}{x_{2}q^{2}w_{V}^{\prime\prime}}[{\vec k}_{\perp}\cdot p_{\perp}^{\prime\prime}+(x_{1}m_{2}+x_{2}m_{1})(x_{1}m_{2}-x_{2}m_{1}^{\prime})]\Bigg\},\\
a_{-}(q^{2}) & = & \frac{N_{c}}{16\pi^{3}}\int dx_{2}d^{2}{\vec k}_{\perp}\frac{h_{{\cal M}}h_{V}^{\prime\prime}}{x_{2}\hat{N}_{1}^{\prime}\hat{N}_{1}^{\prime\prime}}\Bigg\{2(2x_{1}-3)(x_{2}m_{1}+x_{1}m_{2}) -8(m_{1}-m_{2})\Bigg[\frac{p_{\perp}^{\prime2}}{q^{2}}+2\frac{({\vec k}_{\perp}\cdot{\vec q}_{\perp})^{2}}{q^{4}}\Bigg]\nonumber \\
 &  & -[(14-12x_{1})m_{1}-2m_{1}^{\prime}-(8-12x_{1})m_{2}]\frac{{\vec k}_{\perp}\cdot{\vec q}_{\perp}}{q^{2}}\nonumber \\
 &  & +\frac{4}{w_{V}^{\prime\prime}}\Bigg([M^{2}+M^{\prime2}-q^{2}+2(m_{1}-m_{1}^{\prime})(m_{1}-m_{2})](A_{3}^{(2)}+A_{4}^{(2)}-A_{2}^{(1)}) +Z_{2}(3A_{2}^{(1)}-2A_{4}^{(2)}-1)\nonumber \\
 &  & +\frac{1}{2}[x_{1}(q^{2}+q\cdot P)-2M^{2}-2{\vec k}_{\perp}\cdot{\vec q}_{\perp}-2m_{1}(m_{1}^{\prime}+m_{2}) -2m_{2}(m_{1}-m_{2})](A_{1}^{(1)}+A_{2}^{(1)}-1)\nonumber \\
 &  & +q\cdot P\Bigg[\frac{p_{\perp}^{\prime2}}{q^{2}}+\frac{({\vec k}_{\perp}\cdot{\vec q}_{\perp})^{2}}{q^{4}}\Bigg](4A_{2}^{(1)}-3)\Bigg)\Bigg\}.
\end{eqnarray}
\end{widetext}

\section{Helicity Amplitudes}
Here we present the analytical expressions of : $A^i_j, B^i_j, D^i_j$. The helicity amplitudes of $B^*\to B\gamma$ are
\begin{align}
D^{\pm}_0&=\frac{ig_0}{\sqrt{2}}|\vec k|{\rm sin}\theta_{\gamma},\nonumber\\
D^{+}_+&=-D^{-}_-=\frac{ig_0}{2}|\vec k|({\rm cos}\theta_{\gamma}+{\rm cos}2\theta_{\gamma}),\nonumber\\
D^{-}_+&=-D^{+}_-=\frac{ig_0}{2}|\vec k|({\rm cos}\theta_{\gamma}-{\rm cos}2\theta_{\gamma}),
\end{align}
where $|\vec k|$ is the 3-momentum of photon in the rest frame of $B^*$. The helicity amplitudes of $B_s\to B^{(*)}$ are
\begin{align} 
A_0=&\sqrt{\frac{\lambda}{q^2}}f_+,\nonumber\\
A_t=& \frac{1}{2 m_{B_s}^2\sqrt{q^2}}\left\{f_+[m_{B_s}^4-2 m_{B_s}^2 m_R^2+(m_R^2-q^2)^2]\right.\nonumber\\
&\left.f_-[m_{B_s}^4+2 m_{B_s}^2 q^2-(m_R^2-q^2)^2]\right\},\nonumber\\
A^0_0 =&-i \frac{1}{2 m_{R} \sqrt{q^{2}}}\Big[\left(m_{B_s}^{2}-m_{R}^{2}-q^{2}\right)\left(m_{B_s}+m_{R}\right) A_{1}\nonumber\\
&-\frac{\lambda}{m_{B_s}+m_{R}} A_{2}\Big], \nonumber\\
A^+_+ =& i\Big[V \frac{\sqrt{\lambda}}{m_{B_s}+m_{R}}+\left(m_{B_s}+m_{R}\right) A_{1}\Big], \nonumber\\
A^-_- =& i\Big[-V \frac{\sqrt{\lambda}}{m_{B_s}+m_{R}}+\left(m_{B_s}+m_{R}\right) A_{1}\Big], \nonumber\\ 
A^0_t =& -i \frac{\sqrt{\lambda}}{\sqrt{q^{2}}} A_{0},
\end{align}
where $\lambda=\lambda_{B_s}^{(*)},\  m_R=m_{B^{(*)}}$ for $B_s\to B^{(*)}\gamma$,  and $\lambda={\tilde\lambda}_{B_s},\  m_R=\sqrt{p^2}$ for $B_s\to B^{*}(\to B\gamma)l\bar\nu\gamma$. The helicity amplitudes of lepton part read as:
\begin{align} 
B^+_0 & =( \pm i) \cos \theta_{l} \sqrt{q^{2}-m_{l}^{2}} \frac{2 m_{l}}{\sqrt{q^{2}}}, \nonumber\\ 
B^+_+ & =( \pm i)\left(-e^{i \phi}\right) \sin \theta_{l} \sqrt{q^{2}-m_{l}^{2}} \frac{\sqrt{2} m_{l}}{\sqrt{q^{2}}}, \nonumber\\ 
B^+_- & =( \pm i) e^{-i \phi} \sin \theta_{l} \sqrt{q^{2}-m_{l}^{2}} \frac{\sqrt{2} m_{l}}{\sqrt{q^{2}}}, \nonumber\\ 
B^+_t & =( \pm i) \sqrt{q^{2}-m_{l}^{2}} \frac{2 m_{l}}{\sqrt{q^{2}}}, \nonumber\\  
B^-_0 & =( \pm i)\left(-2 \sin \theta_{l}\right) \sqrt{q^{2}-m_{l}^{2}}, \nonumber\\  
B^-_+ & =( \pm i) \sqrt{2} e^{i \phi}\left(1-\cos \theta_{l}\right) \sqrt{q^{2}-m_{l}^{2}}, \nonumber\\ 
B^-_- & =( \pm i) \sqrt{2} e^{-i \phi}\left(1+\cos \theta_{l}\right) \sqrt{q^{2}-m_{l}^{2}}, \nonumber\\  
B^-_t & =0.
\end{align}

\end{appendix}

\end{document}